\documentclass{article}
\usepackage{ijcai24}

\usepackage{comment}
\usepackage{todonotes}
\usepackage{times}
\usepackage{soul}
\usepackage{url}
\usepackage[hidelinks]{hyperref}
\usepackage[utf8]{inputenc}
\usepackage[small]{caption}
\usepackage{graphicx}
\usepackage{amsmath}
\usepackage{amsthm}
\usepackage{booktabs}
\usepackage{algorithm}
\usepackage{algorithmic}
\usepackage[switch]{lineno}
\usepackage{amsfonts}
\usepackage{tabularx}
\usepackage{multirow}
\usepackage{makecell}
\usepackage{enumitem}
\linenumbers

\title{From Factor Models to Deep Learning:\\ Machine Learning in Reshaping Empirical Asset Pricing}
\author{
Junyi Ye$^*$
\and
Bhaskar Goswami$^*$\and
Jingyi Gu$^*$\and
Ajim Uddin\And
Guiling Wang
\affiliations
New Jersey Institute of Technology
\emails
\{jy394, bg362, jg95, ajim.uddin, gwang\}@njit.edu
}
\begin{document}
\nolinenumbers
\maketitle
\def\thefootnote{*}\footnotetext{These authors contributed equally to this work.}\def\thefootnote{\arabic{footnote}}

%\todo[inline]{Everyon, our reference list is very long, we need to remove non important references, write now we have about 3.5 pages of reference, we need to bring it about 2-2.5 pages }

\begin{abstract}
This paper comprehensively reviews the application of machine learning (ML) and AI in finance, specifically in the context of asset pricing.
It starts by summarizing the traditional asset pricing models and examining their limitations in capturing the complexities of financial markets. 
It explores how 1) ML models, including supervised, unsupervised, semi-supervised, and reinforcement learning, provide versatile frameworks to address these complexities, and 2) the incorporation of advanced ML algorithms into traditional financial models enhances return prediction and portfolio optimization. These methods can adapt to changing market dynamics by modeling structural changes and incorporating heterogeneous data sources, such as text and images. In addition, this paper explores challenges in applying ML in asset pricing, addressing the growing demand for explainability in decision-making and mitigating overfitting in complex models. This paper aims to provide insights into novel methodologies showcasing the potential of ML to reshape the future of quantitative finance.
\end{abstract}

\section{Introduction}
 % The finance sector, often hailed as the lubricant of economic society, plays an indispensable role in facilitating smooth economic operations and growth. 
 The finance sector, often recognized as the backbone of economic society, plays an indispensable role in facilitating smooth economic operations and growth. This sector faces unique challenges, stemming from the complexity of financial markets, regulatory constraints, the need for precise decision-making, big data, and the rapid pace of technological change. The integration of Machine Learning (ML) and Artificial Intelligence (AI) into this domain, particularly in asset pricing, is not merely an advancement but a necessity.

Empirical asset pricing models, which aim to elucidate the complex relationships between financial assets and their expected returns, are crucial for investors, fund managers, and policymakers. Traditional models like the Capital Asset Pricing Model (CAPM) \cite{sharpe1964capital} and the Fama-French models \cite{fama2015five} have been foundational yet often struggle to capture the multifaceted and nonlinear dynamics of financial markets. %The predictive accuracy, variable selection, and adaptability of these traditional models are notably challenged by the real-world financial data's complexity \cite{welch2008comprehensive}.

% \todo{dont you think the last sentence of earlier paragraph and first sentence of next para is same?
% }

In literature, the issues of less predictive accuracy, difficult variable selection, and less flexible functional forms of traditional models are well documented \cite{welch2008comprehensive,he2013intermediary}.
[Gu \textit{et al.,} \shortcite{gu2020empirical}, propose addressing these issues by introducing ML in asset pricing.
ML models offer better predictive power and the ability to model complex non-linear relationships with much more flexibility. They also enable the integration of non-traditional data sources, e.g., text, image, video, and audio data, enriching the decision-making processes. 
In addition, fine-tuning and parameter optimization strategies enable continuous learning based on new information, facilitating real-time decision-making. As a result, we see a surge in the application of ML in finance, especially for modeling complexities in asset pricing.

In this paper, we offer a comprehensive review of empirical asset pricing using machine learning (ML). We examine how ML-based approaches have transformed traditional models, providing a renewed perspective on challenges in asset pricing. 
Unlike some previous works that focused solely on either Finance \cite{giglio2022factor} or those detailing the taxonomy of financial-risk tasks linked to machine learning methods (\cite{9249416}),
our approach involves an examination of recent research contributions from both finance and computer science fields, offering a comprehensive view of the interdisciplinary advancements. Moreover, we conduct a critical evaluation of current challenges in adopting ML for empirical asset pricing, identifying research gaps and providing insights into future research directions.
As a result, this review serves as a valuable resource for future researchers, offering a starting point to understand the current advancements in this interdisciplinary domain. Pioneering in its scope, our survey is the first to provide a detailed emphasis on the ML algorithm development perspective specifically for asset pricing, filling a critical void in existing literature.

The organization of the rest of the paper is as follows: 
Section \ref{estimatingriskpremia} addresses the fundamental asset risk premia estimation problem and covers a spectrum from traditional factor models to ML for various tasks.
Section \ref{portfolio_optimization} explores portfolio optimization from traditional approaches to supervised learning and reinforcement learning strategies. 
Section \ref{advanced_techniques} delves into recent advancements in asset pricing techniques.
Section \ref{challengesfuturedirection} examines the challenges and anticipates future trends.

\section{Risk Assessment and Price Prediction}
\label{estimatingriskpremia}
Asset pricing pertains to the valuation of various financial instruments, 
such as equities, options, fixed-income securities, and cryptocurrencies. 
In traditional finance, the fundamental problem of asset pricing boils down to estimating asset risk premia defined as the conditional expected return over the risk-free rate. 

\begin{equation}
    y_{i,t} =  \mathbb{E}[y_{i,t}] + \epsilon_{i,t},
\end{equation}

where $y_{i,t}$ is the individual assets excess return over the risk-free rate, i.e., t-bill rate, with $ \mathbb{E}[\epsilon_{i,t}]=0$. The conditional expected return $\mathbb{E}[y_{i,t}]$ often modeled as an unknown function $g(\cdot)$ of some $P$-dimensional predictors $x_{i,t}$ and parameterized by $\theta$ as: 

\begin{equation}
\label{Eq. Return_es}
    y_{i,t} = g(x_{i,t-1}; \theta) + \epsilon_{i,t}. 
\end{equation}

Asset pricing models commonly incorporate asset-specific characteristics and macroeconomic factors as predictors $X$ for estimating excess return $y$. 

\subsection{Traditional Factor Models}
\label{assetpricingtrad}

Historically, factor models have stood out as pivotal frameworks for asset pricing. The fundamental idea behind these factor-based models is that the high-dimensional predictors $X$ can be replaced with low-dimensional factors $F$. This replacement allows for the explanation of cross-sectional asset returns through their sensitivity to the risk factor. In a factor model, excess return can be modeled as: 

\begin{equation}
\label{eq: conditional}
    y_{i,t} = \alpha_{i,t-1} + \beta'_{i,t-1}f_t + \epsilon_{i,t},
\end{equation}

where, the loading $\beta_{i,t-1}$ represent an asset $i$'s exposure  to a common risk factor $f_t$ and  $\alpha_{i,t-1}$ denotes the mispricing component, assuming $\alpha_{i,t-1} = 0,  \forall  i$ and $t$.

Although some factors, such as industrial production growth, are known and observable, the empirical asset pricing literature assumes that most factors are unobservable or latent. Researchers employ two approaches to estimate these latent factors. The first approach involves constructing characteristic-sorted portfolios based on prior knowledge about the cross-section of returns, and then considering the long-short portfolio returns as the observable factors. The well-known Fama-French Five-Factor model \shortcite{fama2016dissecting}, encompassing market, size, value, profitability, and investment, serves as an example of such factors.
Over the years, the empirical asset pricing literature has reported hundreds of factors that contribute to explaining cross-sectional asset returns. Despite these efforts, empirical evidence indicates that factor models developed using these techniques have failed to produce $\alpha_{i,t-1} = 0$. Researchers attribute this limitation to a partial understanding of the variability in cross-sectional average returns \cite{kelly2019characteristics,giglio2022factor}. Constructing an all-encompassing model in this approach requires complete knowledge of the cross-section of returns and the inclusion of every influencing characteristic. However, our current understanding remains partial at best.

The second approach is to estimate the latent factors from the panel of realized returns using factor analysis techniques, e.g., PCA \cite{connor1986performance}. This approach does not require ex-ante knowledge of the cross-section of average returns. In a static PCA-based approach, the estimation for factors occurs in two steps: firstly, it combines a large set of predictors into linear combinations of latent factors $f$, and then utilizes these latent factors to model excess returns through predictive regression. The relationship between the latent factor $f_t$ and observable characteristics $x_t$ is represented as $x_t = \beta f_t + u_t$. In order to estimate time-varying factors and factor loadings, \cite{kelly2019characteristics} propose IPCA, as:

\begin{equation}
\label{eq: conditional}
    y_{i,t} = x_{i,t-1}\beta_{i}f_t + \epsilon_{i,t}. 
\end{equation}

Here, $\beta\in\mathbb{R}^{N\times K}$ and $f_t\in\mathbb{R}^{K\times T}$ are unknown parameters, and $\epsilon_{i,t}$ is a composite error, including $\alpha_{i,t-1}$. A significant strand of traditional asset pricing literature focuses on developing appropriate techniques to reduce the dimensionality of predictors and estimate these latent factors \cite{feng2020taming,giglio2021asset} (a detailed discussion is provided in Section \ref{sec:Dim_reduction}).

These traditional approaches serve as the building blocks for estimating asset risk premia and the various factors influencing asset returns. However, these methods have multiple limitations. First, the ever-increasing factor universe leads to a large number of free parameters in Equation \ref{eq: conditional}, resulting in inefficient estimations using traditional regression-based models. Second, the factor models rely on a linear approximation of risk exposure based on observable characteristics. Nevertheless, theoretical asset pricing models suggest nonlinearity in return dynamics \cite{bansal2004risks,he2013intermediary}. In the following section, we discuss how integrating machine learning techniques into traditional models can overcome these limitations and further augment their capabilities.

\begin{figure}[t!]
    \centering
    \includegraphics[width=8.3cm]{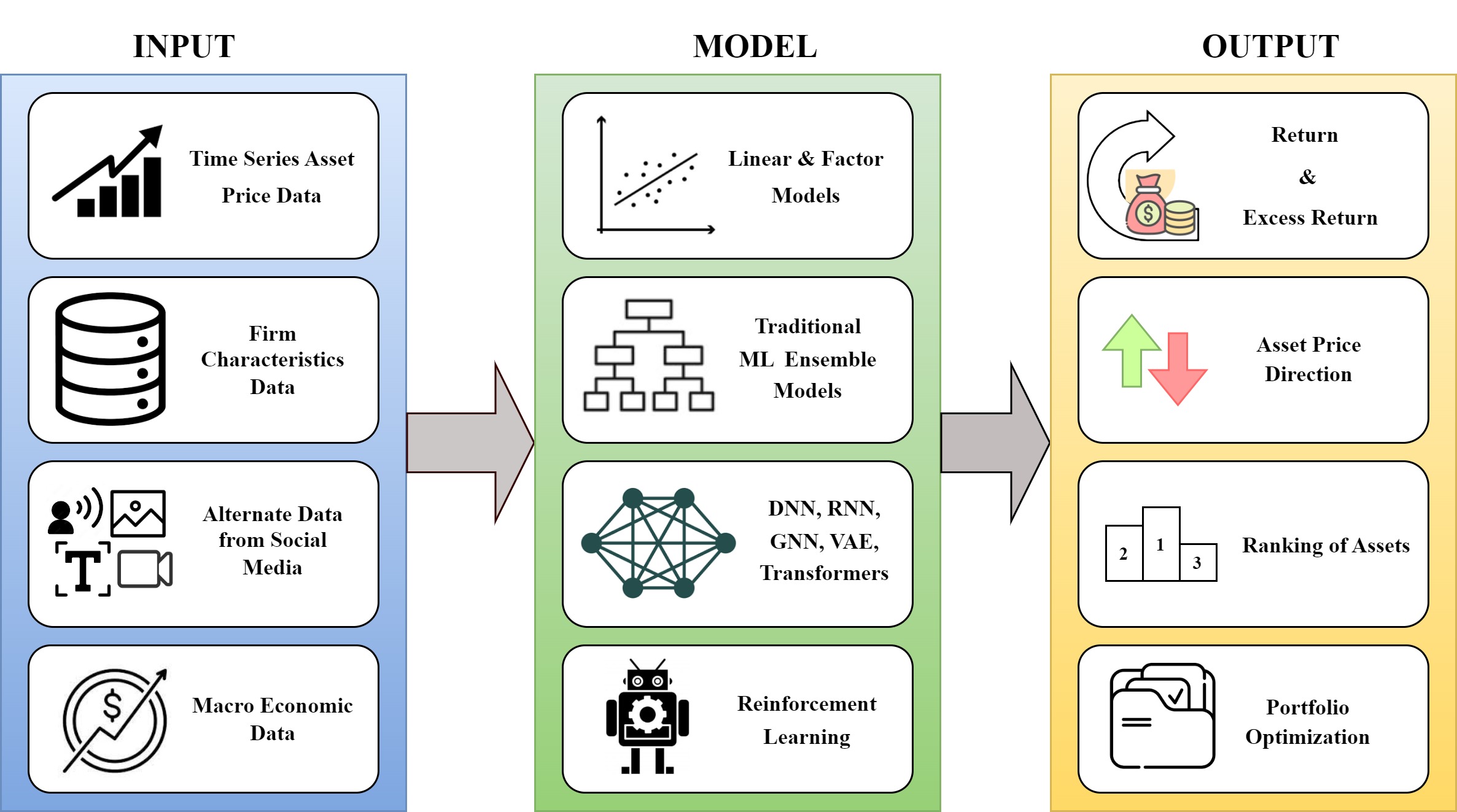}
    \caption{ML in Asset Pricing}
    \label{fig:mlassetpricing}
    \vspace{-0.15in}
\end{figure}

\subsection{AI-Augmented Prediction Models}
\label{MachineLearningModels}

Unlike traditional models, ML takes into account the temporal variability of financial markets and their non-linear interactions. ML also provides built-in solutions for factor selection and handling high-dimensional data. As a result, we observe numerous successful applications of ML models in empirical asset pricing, encompassing risk premia estimation \cite{chen2023deep}, feature engineering \cite{gu2020empirical,long2019deep}, and dimension reduction \cite{kelly2019characteristics,kelly2023modeling}.

To simplify, most ML-based model tries to estimate the function $g(\cdot)$ identified in Equation \ref{Eq. Return_es}. 
The objective is to find the optimal functional representation for the cross-sectional return $y_{t}$ based on the information (characteristics) available at previous time point $x_{i,t-1}$ as:  

\begin{equation}
    \label{eq: ML}
        \hat{g}(\cdot) = \arg \min_{g \in \mathcal{G}} \sum_{i=1}^{N} \sum_{t=1}^{T} (y_{i,t} - g(x_{i,t-1}))^2.
\end{equation}

Researchers have employed various specifications and models with flexible structures to estimate the function $g(\cdot)$. These models range from simple linear models, such as linear regression, to more complex ones like non-linear decision trees, ensemble random forests, gradient-boosted regression trees, and deep neural networks \cite{gu2020empirical,leippold2022machine}. The application domain is also extensive, covering equities \cite{gu2020empirical,chen2023deep}, cryptocurrencies, futures, and options \cite{goudenege2020machine}.

The functional form, as depicted in Equation \ref{eq: ML}, is primarily employed for cross-sectional return prediction. Beyond this, ML applications extend to predicting individual asset returns using time-series models and forecasting price movements or rankings through the use of classification and regression algorithms. %\todo{Junyi: Most papers for asset rankings actually used regression-based models since classification (softmax) does not have an ordinal relationship.; What do you mean by that? Junyi: Should we change to "through the use of classification and regression algorithms" yeah you?}

\begin{figure}[t!]
    \centering
    \includegraphics[width=8.5cm]{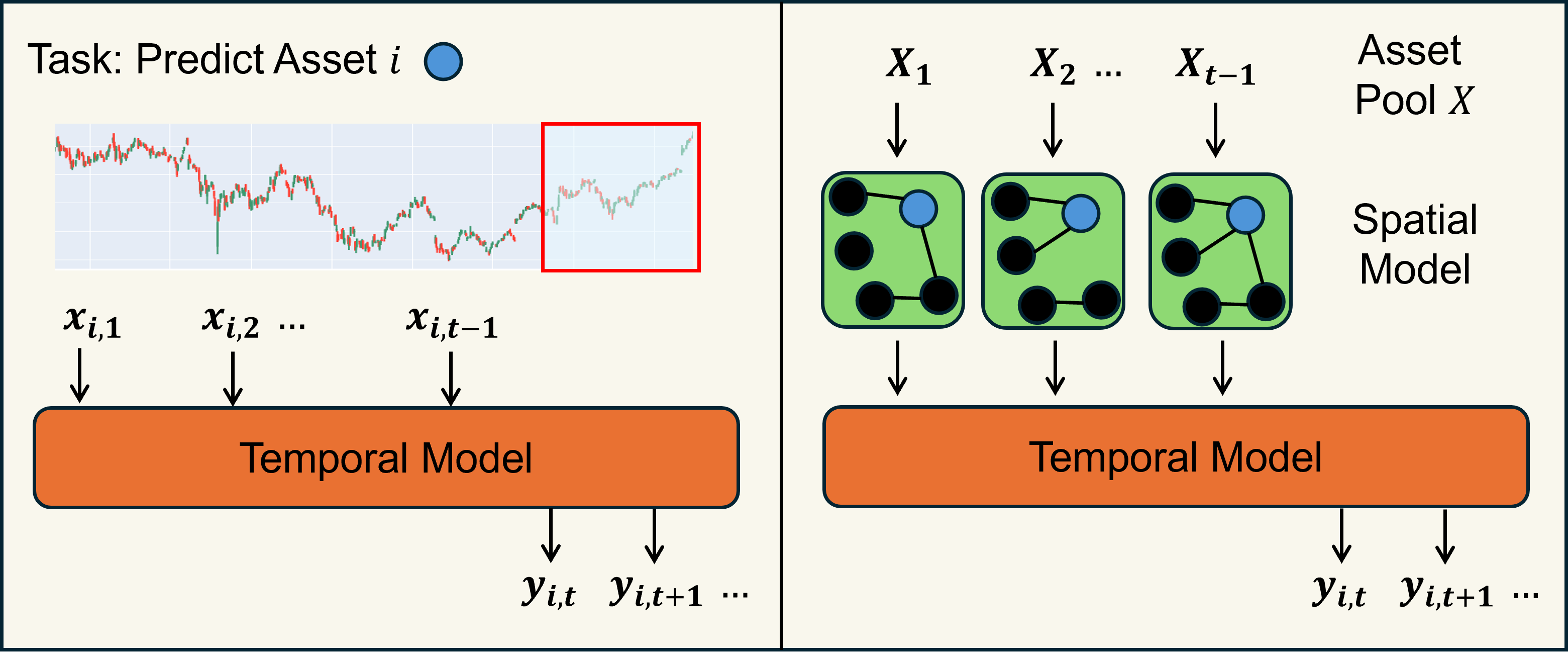}
    \caption{General Pipelines of Temporal Models (Left) and Spatio-Temporal Models (Right).}
    \label{fig:temporal-spatial}
    \vspace{-0.15in}
\end{figure}

Time-series models forecast asset values by analyzing historical price data to identify temporal patterns and dependencies, such as trends, seasonality, and cycles. For return prediction, common evaluation metrics include distance-based measurements such as RMSE, MAE, and MAPE. Asset movement prediction, driven by classification algorithms, focuses on the direction of price changes (upward, stationary, and downward) by examining market trends and correlations. In this context, evaluation metrics such as Accuracy, Precision, Recall, F1 Score, and MCC are employed. Specifically, MCC is often emphasized in financial data analysis due to its effectiveness in handling imbalanced datasets, offering a more nuanced performance measure compared to Accuracy and F1 Score. Meanwhile, asset ranking prediction uses regression models to organize assets within a market or portfolio, aiding investors and fund managers in identifying securities likely to outperform, thereby supporting strategic investment decisions. Ranking predictions are typically evaluated using metrics such as MAP, MRR, and NDCG. Details of these performance measures are presented in Table \ref{tab:metrics}.

In the following sections, we delve into recent advancements in ML models that capture the temporal and spatial dependencies in financial data, highlighting the innovative approaches and techniques that have emerged in prediction and ranking.

\subsubsection{Temporal Models}
To address the inherent temporal dependencies in financial data, a number of temporal models have been developed for individual assets. These models aim to decode historical dynamics to predict future market trends accurately. 
Temporal models operate by harnessing a sequence of data points from prior time stamps, referred to as lagged observations, to project future values. These inputs range from simple constructs, like lagged prices or returns, to more intricate arrays comprising diverse financial indicators and market data. Mathematically, the nexus between historical inputs and future predictions is encapsulated as follows:

\begin{equation}
\label{eq: temporal-models}
y_{i,t}, y_{i,t+1}, \ldots = g_{t}(x_{i,t-1}, x_{i,t-2}, \ldots, x_{i,1}).
\end{equation}

Here, $y_{i,t}, y_{i,t+1}, \ldots$ signify the predicted values for asset $i$ (e.g., future prices or returns) at future time $t,t+1,\ldots$. The terms $x_{i,t-1}, x_{i,t-2}, \ldots, x_{i,1}$ represent the historical input features up to time $t-1$. Each $x_{i,t}$ is a vector that encapsulates one or multiple attributes such as lagged prices, returns, or other pertinent financial indicators at time $t$. The function $g_{t}$ is designed to capture the intrinsic temporal patterns and dependencies within the data, effectively mapping past observations to future values.

The evolution of these models has been significantly influenced by the rise of deep learning, replacing traditional models like ARIMA and VAR.
Starting from the simple feedforward model the focus soon shifted to Recurrent Neural Networks (RNN) and especially Long-Short Term Memory (LSTM) \cite{selvin2017stock}.
Recently, the trend in time-series forecasting has shifted from RNNs towards all-MLP (Multilayer Perceptron) models such as N-BEATS \cite{oreshkin2019n}, which are dominating the time series benchmark leaderboards. Notably, TS-Mixer \cite{ye2023prediction} has outperformed both RNNs and N-BEATS in predicting S\&P 500 returns, showcasing the potential of sophisticated designs, such as residual connections and Mixer blocks in MLP architectures, jointly learning the correlations among features and temporal dynamics.

Another significant effort involves transitioning from single-scale to multi-scale (time, frequency, resolution) approaches. This shift, incorporating different types of scale techniques (Fourier transform, wavelets, downsampling), is crucial for dissecting the multifaceted behavior of stocks, ranging from rapid intraday fluctuations to long-term trend formations. MTDNN \cite{liu2020multi} utilizes wavelet-based and downsampling techniques to analyze stock patterns at varying scales, from fine-grained to broader trends. Transformer-based approach \cite{ding2020hierarchical} with its Multi-Scale Gaussian Prior
excels in capturing long-term, short-term as well as hierarchical dependencies of financial time series.

\begin{table}[t!] % need to mention the table in the paper
    \centering
    \tiny
    \renewcommand{\arraystretch}{1.5}
    \begin{tabular}{|m{0.02\textwidth}<{\centering}|m{0.13\textwidth}<{\centering}|m{0.26\textwidth}<{\centering}|}
    \hline
         \textbf{Task} & \textbf{Evaluation Metric}&  \textbf{Formula}\\
         \hline
         \multirow{3}{=}{\rotatebox[origin=c]{90}{\parbox{1.3cm}{\centering Return\\Prediction}}}
         & RMSE ($\downarrow$)& $\sqrt{\frac{1}{T} \frac{1}{N}\sum_{(i,t)\in \Omega}(y_{i, t}-\hat{y}_{i, t})^2}$ \\ 
         & MAE ($\downarrow$)& $\frac{1}{T} \frac{1}{N}\sum_{(i,t)\in \Omega}|y_{i, t}-\hat{y}_{i, t}|$ \\
         & MAPE ($\downarrow$)& $\frac{1}{T} \frac{1}{N} \sum_{(i,t) \in \Omega} \left| \frac{y_{i, t} - \hat{y}_{i, t}}{y_{i, t}} \right| \times 100\%$ \\
         \hline
         \multirow{3}{=}{\rotatebox[origin=c]{90}{\parbox{1.9cm}{\centering Movement \\ Prediction}}}
         & Accuracy ($\uparrow$)& $\frac{\sum_{i=1}^{N} I(\hat{D}_i = D_i)}{N}$ \\
         & Precision ($\uparrow$)& $\frac{\text{TP}}{\text{TP+FP}}$ \\
         & Recall ($\uparrow$)& $\frac{\text{TP}}{\text{TP+FN}}$ \\
         & F1 Score ($\uparrow$)& $\frac{2 \cdot \text{Precision} \cdot \text{Recall}}{\text{Precision} + \text{Recall}}$ \\
         & MCC ($\uparrow$)& $ \frac{TP \times TN - FP \times FN}{\sqrt{(TP + FP)(TP + FN)(TN + FP)(TN + FN)}}$\\
         \hline
         \multirow{3}{=}{\rotatebox[origin=c]{90}{\parbox{1.2cm}{\centering Ranking \\ Prediction}}}
         & MAP ($\uparrow$)& $\frac{1}{T} \frac{1}{K}\sum_{(k,t)} P(k)_t \cdot \text{rel}(k)_t$ \\
         & MRR ($\uparrow$)& $\frac{1}{T} \sum_{t=1}^{T} \frac{1}{rank_{t}}$ \\
         & NDCG ($\uparrow$)& $\frac{\text{DCG}}{\text{IDCG}}, \text{ where } \text{DCG} = \sum_{k=1}^{K}  \frac{\text{rel}(k)}{\log_2(k+1)}$\\
         \hline
         \multirow{6}{=}{\rotatebox[origin=c]{90}{\parbox{2.9cm}{\centering Portfolio \\ Optimization}}}
         & Accumulated Return ($\uparrow$)& $\prod\limits_{t=1}^T\sum_i^N w_{i,t-1} y_{i,t}$ \\
        & Volatility ($\downarrow$)& $\sigma(PR_t)$ \\
        %$\sqrt{\frac{1}{T} \frac{1}{N} \sum_{(i,t) \in \Omega} (\ln(1+y_{i,t}) - {\ln(1 + \Bar{y}_{i,t}}))^2}$ \\
         & Maximum Drawdown ($\uparrow$)&  $\max_{0\leq a \leq b \leq T} 1-\frac{\prod\limits_{t=1}^b\sum_i^N w_{i,t-1} y_{i,t}} {\prod\limits_{t=1}^a\sum_i^N w_{i,t-1} y_{i,t}} $\\
         & Sharpe Ratio ($\uparrow$)&  $\frac{\sum_{t=1}^T(PR_{t})/T}{\sigma(PR_t)}$ \\
         & Calmar Ratio ($\uparrow$)& $\frac{\sum_{t=1}^T(PR_{t})/T}{\text{MDD}}$ \\
         & Sortino Ratio ($\uparrow$)& $\frac{\sum_{t=1}^T(PR_{t})/T}{\text{DD}}$ \\
         \hline
    \end{tabular}
    \caption{Evaluation Metrics Used in Empirical Asset Pricing\protect\footnotemark. $\uparrow$ means the larger the better while $\downarrow$ means the smaller the better.}
    \vspace{-0.15in}
    \label{tab:metrics}
\end{table}

\addtocounter{footnote}{0}
\footnotetext[\thefootnote]{\footnotesize \textbf{Note:} RMSE: Root Mean Square Error, MAE: Mean Absolute Error, MAPE: Mean Absolute Percentage Error, MCC: Mathews Correlation Coefficient, MAP: Mean Average Precision, MRR: Mean Reciprocal Rank, NDCG: Normalized Discounted Cumulative Gain, $T$: Number of timestamps, $N$: Number of assets, 
% $\mathbf{y}$: Excess Return , 
$D$: Price Movement Direction, TP: True Positive, TN: True Negative, FP: False Positive, FN: False Negative, $K$: Maximum top-ranked assets evaluated, %the maximum number of top-ranked assets considered in the evaluation, 
$P(k)$: Precision at Rank $k$, %the precision measures the proportion of the model's top-$k$ ranked assets that are deemed optimally suitable by the ground truth, 
$\text{rel}(k)$: 1 if rank $k$ asset is suitable, 0 otherwise, %a binary relevance indicator that equals 1 if the asset at rank $k$ by model is considered optimally suitable, and 0 otherwise, 
$rank$: Position of first relevant asset, % (i.e. the top-ranked asset by model that is considered optimally suitable) at time $t$, 
DCG: Discounted Cumulative Gain, % that measures the total asset relevance, 
IDCG: Ideal Discounted Cumulative Gain, %, a normalization factor used to compare the DCG score to the best possible (ideal) DCG score, 
$\sigma(\cdot)$: Standard Deviation, PR: Portfolio Return, DD: Downside Deviation, MDD: Maximum Drawdown.}
%\todo{Junyi: I have updated the evaluation metrics for ranking prediction. Most of them are from recommendation system and they are complex (difficult to explain with single sentence for asset pricing). I made some explanation in the footnote. Please feel free to modify them or let me know if there is any questions.}
% \todo{i see, but yiou probaby need to make it compact, can you try removing some details on those explanation, because note is almost half page. }

% MAP and NDGG:https://dl.acm.org/doi/pdf/10.1145/3292500.3330663
% MAP example: https://www.evidentlyai.com/ranking-metrics/mean-average-precision-map#:~:text=Mean%20Average%20Precision%20(MAP)%20at,relevant%20items%20at%20the%20top.
% MRR: https://dl.acm.org/doi/pdf/10.1145/3309547
\subsubsection{Spatio-Temporal Models}
The financial market exhibits interconnectivity, where fluctuations in the price of one asset can influence the pricing dynamics of its affiliated counterparts.
Temporal models adeptly capture chronological data but often neglect spatial interconnections, which are pivotal in the financial market. Incorporating these spatial connections can significantly enhance prediction accuracy and reduce noise in asset pricing. Many studies have introduced Graph Neural Networks (GNNs) to model the intricate relationships between assets, capturing spatial dynamics in the process. GNNs are particularly adept at learning sparse spatial dependencies, a trait that aligns well with the complex and nuanced nature of financial markets.

The general form of spatio-temporal is denoted as:

\begin{equation}
\label{eq: spatio-temporal-model}
y_{t}, y_{t+1}, \ldots = g_t[g_s(X_{t-1}), g_s(X_{t-2}), \ldots, g_s(X_{1})].
\end{equation}

Here, $X_{t}, X_{t-1}, \ldots, X_{1}$ represent the feature matrices of the asset pool (e.g., a collection of assets) at past time stamps, and $y_{t}, y_{t+1}, \ldots $ are the predicted values for the future time steps. Most existing spatio-temporal models first capture spatial correlations with $g_{s}$, such as GNN, among assets and then their temporal evolutions with temporal models $g_{t}$.

To model spatial dynamics, researchers adopt diverse relational frameworks for constructing graphs, capturing the unique similarities among entities in the financial market. 
\cite{chen2018incorporating} explored shareholding ratios to map corporate interconnections, employing graph embedding techniques such as DeepWalk, LINE, node2vec, and Graph Convolutional Network (GCN). 
Meanwhile, HATS \cite{kim2019hats} applied Graph Attention Networks (GAT) to analyze graphs based on ownership and organizational structures. 
Addressing the limitations of traditional GNNs, which are restricted by pre-defined firm relations
AD-GAT \cite{cheng2021modeling} introduces an attribute-mattered aggregator for capturing attribute-sensitive momentum spillovers. 
RSR \cite{feng2019temporal} employed Temporal Graph Convolution, they use sector similarity and supply-chain network to analyze relational dynamics and rank stocks by their revenue potential. 
Similarly, STHAN-SR \cite{sawhney2021stock} enhances stock selection for quantitative trading through a Spatio-Temporal Hypergraph Attention Network. This model utilizes a neural hypergraph structure, with hyperedges representing industry and corporate connections.
\cite{uddin2021attention} utilized attention mechanisms to develop evolving networks, linking firms through Pearson Correlation, and learned from positive and negative graphs separately.
Besides GNNs, DTML \cite{yoo2021accurate} uses a Transformer based model to encode asymmetric, and dynamic stock correlations.

These spatial models integrate temporal recurrent neural networks such as LSTM and GRU and evolve into spatio-temporal models. This integration results in a more sophisticated and accurate representation of market dynamics, effectively capturing both spatial connections and temporal evolutions within the financial landscape.

\section{Portfolio Optimization}
% written by Jingyi

\label{portfolio_optimization}
%\subsection{Traditional Financial Approaches}
Portfolio optimization is a crucial component of asset pricing. Investors construct an investment portfolio comprising diverse assets including stocks, foreign currency, futures, bonds, cryptocurrencies, and other securities.  
The Modern Portfolio Theory (MPT)\cite{mpt} revolutionized the landscape of portfolio construction and risk management. MPT argues that investors, being inherently risk-averse, can strategically design portfolios to either maximize expected returns within a specified level of risk or, conversely, to minimize risk for a given level of expected return. MPT emphasizes the significance of diversification, suggesting that by spreading investment across various assets without positive correlation, investors can achieve a more optimal risk-return trade-off. %\todo{i think its better if we remove traditional approach}
The utility maximizing portfolio solution can be defined as: 
\begin{equation}
    w^* = \frac{1}{\gamma} \Sigma^{-1} \mu, 
\end{equation}
where, $\mu \in\mathbb{R}^N$ and $\Sigma\in\mathbb{R}^{N\times N} $ represent the mean and variance of return on $N$ assets, and $\gamma$ represent investros risk aversion. An investor can develop an optimal portfolio of $w^*\in\mathbb{R}^N$ in two ways. First, by using models to estimate the return distribution and treating the estimations as known parameters $\mu$ and $\Sigma$, then selecting weights $w$ that optimize the utility. Second, by utilizing models to optimize the portfolio through parameterizing the weights. Here, we discuss how ML facilitates both of these methods.

\subsection{Supervised Learning Techniques}
%\todo{Jingyi: 3.2 is restructured. Instead of listing works, we explain how section2 model and return results are used to construct portfolio using supervised models, which is different from directly determining weights in RL}
%With the development of machine learning, 
Supervised Learning models can facilitate portfolio optimization by estimating $\mu$ and $\Sigma$ of assets. As discussed in Section \ref{estimatingriskpremia}, ML can predict returns, rank, and movements. Investors can choose appropriate assets based on these predictions and develop a long-short portfolio. 
Firstly in the process of choosing assets: (1) Given a return prediction model or rank prediction model, investors can take a long position on assets with the highest predicted future returns or highest ranking, and a short position on assets with the lowest predicted future returns or lowest ranking \cite{uddin2021attention,lin2021learning,duan2022factorvae}. (2) Given a movement prediction model, the portfolio can be developed by taking a long (short) position in an asset with a predicted upward (down) direction \cite{yoo2021accurate}. Subsequently, during the portfolio construction process, investors establish position at the beginning of period $t-1$ and liquidate/rebalance the portfolio at $t$. Portfolio are typically allocated with equal weights $w_{i,t} = \frac{1}{N}$ or weighted weights $w_{i,t} = \frac{v_{i,t}}{\sum_{i=1}^N{v_{i,t}}}$.

\subsection{Reinforcement Learning Methods}
\begin{figure}[t!]
    \centering
    \includegraphics[width=5cm]{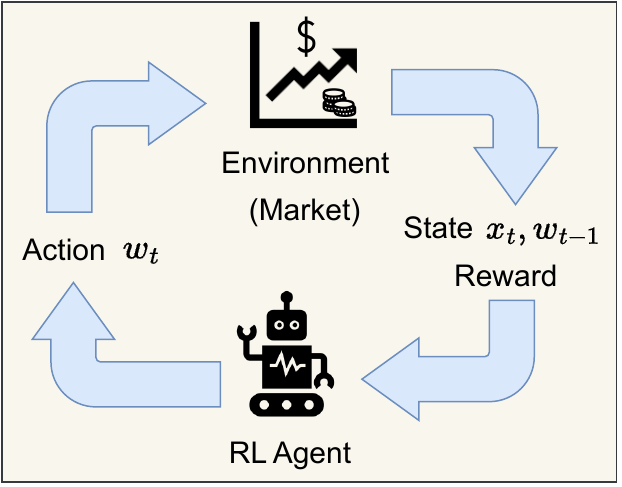}
    \caption{General Framework for RL in Portfolio Optimization} %at time $t$. $x_t$ is the price/return vector, $w_{t-1}$ and $w_t$ are previous and current portfolio weight vectors respectively.}
    \label{fig:rl}
    \vspace{-0.14in}
\end{figure}

The portfolio weight $w_{i,t}$ among $N$ asset at time $t$ can also be directly determined with the aim of maximizing return and minimizing risk. 
The most prominent approach to this end is Reinforcement learning (RL). Leveraging its nature of sequence decision-making RL utilizes neural networks to approximate the action policy function and estimate the reward function. In the RL environment as shown in Figure \ref{fig:rl}, the state $s_t$ at the beginning of each period is defined as the pair of historical prices $x_t$ and previous portfolio weight $w_{t-1}$, observable in the environment; deep neural networks are employed to determine the portfolio weight vector $w_t$ as the action output $a_t$; the agent, representing the portfolio manager, performs trading actions and obtains periodic logarithmic returns as the reward $r_t$. The RL agent aims to find an optimal policy to maximize the action-value function $Q^\pi (s,a)$:
\begin{gather}
    Q^\pi (s,a)=\mathbb{E}[\sum_{i=t}^{\infty}\gamma^i r_{t+i}|s_t=x_t,a_t=w_{t-1}]
\end{gather}
where $\gamma_i$ is the discount factor. A notable contribution in this domain is EIIE ~\cite{jiang2017deep}, which proposes a financial-model-free RL approach with deep learning solutions for the portfolio optimization problem. It assumes \emph{Zero market impact}, positing that trading activities do not significantly affect market prices due to relatively small investment capital. EIIE is trained in an Online Stochastic Batch Learning (OSML) scheme and conducts three distinct model species: CNN, RNN, and LSTM. This framework is implemented across various asset classes, including stocks and cryptocurrencies.

Following EIIE, a number of recent studies utilize deep RL for portfolio optimization under more realistic trading scenarios. In \cite{wang2019alphastock,xu2021relation}, attention mechanisms are employed to model interrelationships among assets. In these studies, short sales are incorporated under simple assumptions, such as investors can borrow without limitations and will hold equal long-short positions. Some other works focus on optimizing the reward function with Imitative Recurrent Deterministic Policy Gradient \cite{liu2020adaptive}, and contrastive learning \cite{lien2023contrastive}. Trading execution and trading costs are also considered. The Hierarchical RL framework developed in \cite{wang2021commission} enables the placement of small-sized limit orders at desired prices and quantities during the trading period to minimize trading costs. Smart Trader~\cite{yang2022smart} employs a Geometric Brownian Motion process for portfolio weight and optimal trading point decisions to achieve excess intraday profit. Deep RL also contributes significantly to algorithmic trading. For example, iRDPG  applied Imitative Learning and Behavior Cloning to create a trading bot that consistently takes intraday greedy actions, with the aim of developing a profitable policy \cite{liu2020adaptive}. Meta Trader initially acquires multiple diverse trading policies and subsequently learns to choose the most suitable policy based on market conditions \cite{niu2022metatrader}.

Additionally, a notable research avenue focuses on the generalized RL framework. The openly accessible FinRL library ~\cite{liu2022finrl} provides diverse market environments with data APIs and automatic pipelines. It enables users to select popular benchmarks based on their requirements and design novel strategies. Margin Trader \cite{gu2023margin} integrates margin accounts and constraints to enhance the realism of the trading environment for both long and short positions.

\section{Innovations in Asset Pricing Techniques}
\label{advanced_techniques}

In recent years, the impact of ML on finance has been most prominently recognized in its role in price prediction and portfolio optimization. However, the scope of ML's contributions extends far beyond these domains. Notably, ML has made significant strides in the development of advanced techniques and processes for asset pricing. These innovations not only enhance the accuracy of predictions but also simplify the application of traditional models. This section explores the multifaceted contributions of ML, specifically in advancing techniques that facilitate asset pricing studies.

\subsection{Dimensionality Reduction}
\label{sec:Dim_reduction}

In asset pricing, researchers are often faced with a `factor zoo' – a term referring to the extensive array of factors (hundreds, sometimes) proposed in the literature for predicting asset returns. This not only complicates the models but also raises concerns about overfitting and interpretability. This is where dimensionality reduction techniques become invaluable. The primary motivation for employing dimensionality reduction in financial data analysis is to distill this multitude of factors into a more manageable, representative set. In traditional finance, Partial Least Squares (PLS) \cite{gu2020empirical} and Principal Component Analysis (PCA) \cite{giglio2021asset}, along with its advanced variants, are widely used to achieve these objectives. Factor selection models, like LASSO, is another popular alternative. For example, Feng et al. \shortcite{feng2020taming} proposed a two-step LASSO approach to identify the most influential factors among the 150 previously identified factors for explaining cross-sectional returns.

In recent years, autocoder-based models have gained significant traction for dimension reduction in financial data. In \cite{gu2021autoencoder,uddin2020latent}, the authors use auto-encoder to learn latent asset pricing factors by jointly modeling both asset pricing characteristics and excess return.  
Building on this, Variational Autoencoders (VAEs) advance the concept by introducing a probabilistic twist to the encoding process, effectively distilling meaningful factors from noisy market data.
% Besides reducing data complexity, \textcolor{red}{VAE is utilized to distill effective factors from noisy market data [this sentence seems randomly poped up?I added a sentence to motivate, you can add another connecting sentence]}. 
FactorVAE \cite{duan2022factorvae} addresses the challenge of low signal-to-noise ratios in financial datasets by fusing dynamic factor models (DFM) with a VAE framework, thereby harnessing factors as latent variables to enhance asset pricing models. 
Concurrently, DiffusionVAE (D-Va) \cite{koa2023diffusion} combines a hierarchical VAE structure with diffusion techniques to efficiently process stochastic stock data, ensuring precise, denoised predictions.
%\todo{you can shorten the discussion on D-va. Junyi: I added a connecting sentence between Autoencoder and VAE. Also shorten the D-va; thanks }

\subsection{Imputing Missing Data}
Financial data, especially in asset pricing, often involves missing values.  These values may be random or non-random and can occur for multiple reasons, including poor data curation or intentional non-reporting. In dealing with missing values, historically, researchers and practitioners have either removed observations with missing values or imputed the missing values with zeros or with class mean \cite{bryzgalova2022missing}. However, both approaches have major drawbacks, as they may either remove important information or alter the distribution of the data.

The advancement of recommender systems offers a new avenue for tackling financial missing data. In \cite{uddin2022missing}, a coupled matrix factorization approach is proposed to impute missing analyst earnings forecasts. In their work, the authors augment missing analyst data with firm characteristics data and then use the imputed values for firm return prediction. In \cite{beckmeyer2022recovering}, the authors use a transformer to impute missing firm characteristics data. Additionally, tensor imputation shows significant promise for imputing spatio-temporal data in finance \cite{uddin2022machine,zhou2023fast}.
Despite these efforts, research in this area is still limited. Advanced DL methods can efficiently impute missing values by incorporating non-linearity and spatio-temporal interactions in financial data, thereby offering an exciting direction for future research.

\subsection{Incorporating Alternate Data}
The fusion of alternate data streams—text, images, and speech—with traditional financial information is revolutionizing asset pricing. This surge in popularity is significantly driven by the advancements in transformers and LLMs. By harnessing the power of Computer Vision (CV) and Natural Language Processing (NLP), this multimodal approach augments traditional financial analyses, providing a richer, more nuanced view of market dynamics. The result is the creation of pricing models that are not only more accurate but also significantly more sophisticated.

In financial technology, incorporating image analysis enhances data processing significantly. Traders often rely on visual cues from financial charts, a concept advanced by \cite{cohen2020trading}, who reimagined time-series analysis as an image classification task using models like CNNs. Building on this, \cite{zeng2021deep} transformed price data of multiple assets into 2D images, utilizing video prediction algorithms for market dynamics analysis. 

Textual data, encompassing stock descriptions and social media content, has emerged as a critical component in investment recommendation and market analysis frameworks. Advanced methodologies, such as the Time-Aware Graph Relational Attention Network (TRAN) \cite{ying2020time} %and the approaches delineated in \cite{wang2021heterogeneous}, 
demonstrate the efficacy of integrating textual content with financial datasets to augment stock profiling and recommendation systems. In particular, models like %MAN-SF \cite{wang2021heterogeneous} and the 
Prediction-Explanation Network (PEN) \cite{li2023pen} underscores the paramount importance of synergizing social media inputs, inter-stock relational dynamics, and traditional financial indicators, placing a pronounced emphasis on the  sentiments analysis for improving predictive accuracy in stock movements.

Furthermore, the pioneering VolTAGE model \cite{sawhney2020voltage} represents a significant stride in the realm of multimodal data integration for stock volatility prediction. This model ingeniously incorporates vocal cues from corporate executives' earnings calls, analyzing them in conjunction with traditional financial data within a Graph Convolutional Network (GCN). The deployment of an inter-modal, multi-utterance attention mechanism within VolTAGE is particularly noteworthy, exemplifying a comprehensive approach to data synthesis. %This methodology effectively harnesses both verbal and vocal cues, distilling a wealth of nuanced information into actionable insights, thereby offering a holistic perspective on the multifaceted dynamics of stock volatility prediction.

\subsection{Denoising and Non-IID Adaptation}
Recent trends in financial data prediction have highlighted the integration of cutting-edge AI techniques.% to improve accuracy and profitability. 
Among these, contrastive learning and Mixture of Experts (MoE) stand out for their effectiveness. 

Contrastive learning, initially finding significant success in computer vision tasks, has been adapted to refine models' predictive capabilities in financial domains by contrasting positive (similar) and negative (dissimilar) sample pairs. This approach is exemplified in techniques such as Copula-based Contrastive Predictive Coding (Co-CPC) \cite{wang2021coupling} and the Contrastive Multi-Granularity Learning Framework (CMLF) \cite{hou2021stock}. Co-CPC excels in filtering out noise and enhancing stock representations by contrasting data from consecutive time points, while CMLF uses its dual contrastive approach to tackle the intricacies of data granularity and temporal relationships, thereby improving prediction accuracy.

Currently, MoE is a leading-edge architecture in Large Language Models (LLMs) that significantly enhances predictive modeling. Researchers utilize it to address complexities often overlooked by the conventional i.i.d. (independent and identically distributed) assumption in financial data. MoE leverages a set of specialized sub-models or `experts', each adept at interpreting specific data patterns or market conditions, countering the traditional one-size-fits-all approach. A sophisticated gating network, termed `router', orchestrates the outputs of these experts, ensuring that the most relevant insights are prioritized based on the current data, thus dynamically adapting to market heterogeneity. Advanced implementations of MoE, such as the Temporal Routing Adaptor (TRA) \cite{lin2021learning} and the Pattern Adaptive Specialist Network (PASN) \cite{huang2022pattern}, underscore its effectiveness. TRA optimizes the assignment of data to predictors, facilitating accurate recognition of temporal patterns, while PASN's Pattern Adaptive Training fosters autonomous adaptation to new and evolving market dynamics. Collectively, these enhancements fortify the robustness and precision of financial market predictions, surpassing the limitations of traditional i.i.d.-based predictive models.

\section{Challenges \& Future Direction}
\label{challengesfuturedirection}
Recent developments have made strides in addressing complex issues within asset pricing, yet a substantial array of challenges demands further attention. These include overarching challenges like overfitting and model interpretability, along with domain-specific issues such as managing noisy data and complying with regulatory standards. In this section, we critically examine these challenges, offering insights into potential research initiatives for effective resolution.

\begin{itemize}[leftmargin=*]
    \item \textbf{Data Availability and Quality:} Financial data are often proprietary, with limited access for academic researchers. Some data are provided by third-party vendors (e.g., WRDS, IBES) at a significant cost and with notable time delays. Consequently, existing research predominantly focuses on indexes and the equity market based on historical data, neglecting other markets like bonds and derivatives. Additionally, unlike other domains, there is a lack of unified test data for evaluating asset pricing models. Establishing a comprehensive dataset with unified assets or synthetic data that reflects the complexity of the financial market is essential for robust and equitable model evaluation.

    \item \textbf{Market Dynamics and Structural Changes:}
    Financial data demonstrates time-varying dynamics and structural changes due to economic uncertainty, geopolitical events, and investor behavioral assumptions. Additionally, the no-arbitrage theory suggests that if a model can identify any mispricing, it will evaporate quickly, rendering the model absolute. As a result, it is challenging for deep learning models to adapt to these changes. Although advancements in NLP facilitate the incorporation of news and sentiment into asset pricing models, the real-time integration of this information into the model is still in its infancy. Modeling structural changes, with online learning and meta-learning can be a step forward in modeling this ever-evolving financial landscape.
    
    \item \textbf{Model Complexity and Overfitting Risks:} Another issue with the time-varying dynamics of financial data is that complex models, e.g., Transformers and GNNs, often perform well on training data but poorly on unseen data. One interesting future research direction involves using algorithms such as Meta-learning, one-shot learning, and ensemble learning, along with early stopping and dropout techniques, to develop more generalized asset pricing models.

    \item \textbf{Regulatory Compliance:} The financial industry is heavily regulated, with strict requirements demanding transparency, accountability, and adherence to established guidelines. Ensuring that AI algorithms comply with regulations such as GDPR, MiFID II, and Basel III entails documentation, robust validation, and ongoing monitoring. Collaborative efforts between financial institutions, AI developers, and regulatory authorities can facilitate the implementation of ethical AI principles and foster transparency in the AI-driven financial landscape.
    
    \item \textbf{Model Interpretability and Fairness:} Deep learning models are infamous for their 'black box' nature and algorithmic bias. In contrast, both regulators and investors are more concerned about the explainability and fairness of a model than predictive accuracy. Understanding the causal inference and economic significance of a model allows them to make more informed decisions and design better policy alternatives. The rise of explainable AI represents a significant leap in this direction. An exciting avenue for future research could involve identifying the balance between interpretability and performance, offering greater transparency to financial models.
%\todo{should we merge last two?}
\end{itemize}

Apart from the aforementioned challenges, the problem of missing values, nonlinear data structure, and heterogeneous data sources persist in finance.  Despite some efforts to address these challenges, they present compelling opportunities for future research.

\section{Conclusion}
\label{conclusion}
In this paper, we aim to encapsulate the ever-growing literature on financial AI. Our motivation is to offer researchers and practitioners an insight into ML as a versatile tool for modeling the intricate workings of financial markets. While much of the current research in empirical asset pricing using ML focuses on enhancing prediction performance, this review demonstrates that such applications represent only a fraction of ML's potential in finance. In less explored areas such as model explainability, optimization, data augmentation, and compliance monitoring, AI can make valuable contributions in the near future.

\bibliographystyle{named}
\bibliography{ijcai24}

\end{document}